\documentclass[fleqn,usenatbib]{mnras}

\usepackage{newtxtext,newtxmath}
\usepackage[T1]{fontenc}

\DeclareRobustCommand{\VAN}[3]{#2}
\let\VANthebibliography\thebibliography
\def\thebibliography{\DeclareRobustCommand{\VAN}[3]{##3}\VANthebibliography}

\usepackage{graphicx}	% Including figure files
\usepackage{amsmath}	% Advanced maths commands
\let\oldAA\AA
\renewcommand{\AA}{\text{\normalfont\oldAA}}
\title[A LOFAR view of 3C\,294]{The high-redshift radio galaxy 3C\,294 at low frequencies: radio detection of the X-ray Ghost}

\author[V. H. Mahatma]{
V.\,H.\,Mahatma,$^{1,2}$\thanks{E-mail: v.mahatma93@gmail.com}
A.\,C.\,Fabian,$^{3}$ and L.\,K.\,Morabito$^{4,5}$
\\
$^{1}$Cavendish Laboratory, University of Cambridge, 19 JJ Thomson Avenue, Cambridge CB3 0HE, UK\\
$^{2}$ Kavli Institute for Cosmology, University of Cambridge, Madingley Road, Cambridge CB3 0HA, UK\\
$^{3}$Institute of Astronomy, University of Cambridge, Madingley Road, Cambridge CB3 0HA, UK\\
$^{4}$Centre for Extragalactic Astronomy, Department of Physics, Durham University, South Road, Durham DH1 3LE, UK\\ 
$^{5}$Institute for Computational Cosmology, Department of Physics, Durham University, South Road, Durham DH1 3LE, UK
}

\date{Accepted 2025 December 18. Received 2025 December 16; in original form 2025 September 25}

\pubyear{2023}

\begin{document}
\label{firstpage}
\pagerange{\pageref{firstpage}--\pageref{lastpage}}
\maketitle

\begin{abstract}
We report on the very first radio detection associated with the peculiar hourglass-morphology X-rays surrounding 3C \,294 at $z=1.8$. Using International Low Frequency Array (LOFAR) data at 144 \,MHz and \textit{Chandra} data at 0.3-6 \,keV, we find that the co-spatial diffuse radio and X-ray emission is well described by synchrotron and inverse-Compton processes by the same electron population. Through modelling of this rare low-energy plasma, we find that the most defining property of the electrons up-scattering CMB photons at this redshift is very low electron Lorentz factors ($\gamma_{\text{max}}\ll 10^{4}$ and $\gamma_{\text{break}}\lesssim 10^{3}$) in the lobe: deep low frequency ($<150$\,MHz) observations are critical to the detection of radio lobes at high redshift. The physical conditions imply a total energy in the diffuse emission significantly greater than that implied by the temperature of the protocluster gas: 3C \,294 is one of the most powerful known radio-loud systems in a dense protocluster environment. Through resolved spectral analysis of archival radio data up to 15\,GHz, we find evidence that the inner hotspots are due to restarted activity, while the outer hotspots remain energetic, suggesting a rapid duty cycle while the jet precesses. This allowed the low-energy aged plasma driving the X-rays to remain spatially distinct from the high-energy plasma. Together, our results promise a revelation of AGN-related radio emission at high redshift using future low-frequency arrays such as SKA-LOW.
\end{abstract}


%%%%%%%%%%%%%%%%%%%%%%%%%%%%%%%%%%%%%%%%%%%%%%%%%%

%%%%%%%%%%%%%%%%% BODY OF PAPER %%%%%%%%%%%%%%%%%%

\section{Introduction}
The energetic plasma in radio galaxies gives important information on high-energy physics: the expanding lobes driven by continuous jet activity contain high-energy ($\gamma>1$) particles that are cooling through the synchrotron and inverse-Compton processes. Understanding these processes is important, as radio galaxy lobes (which contain the bulk of the energetic material) are known to be in direct contact with their environments \citep[see review by][]{worr02}. They drive shocks and excavate cavities in hot gas located in galaxy groups and clusters \citep{cros11,kraf12,vags19}, giving viable solutions to the cooling catastrophe and the steep decline of the galaxy mass function at higher stellar masses \citep{fabi12,morg17}. Knowledge of the physical conditions in the lobes and of the volume fraction they occupy in the Universe over cosmic time drives a better understanding of their effects on their environments. The role of these processes (collectively known as AGN feedback) cosmologically is yet to be understood.

At high redshifts ($z\geqslant1$), the radio galaxy population also trace dense environments such as protoclusters \citep{ramo13,hatc14}, commonly defined as a region of volume which will collapse into a $M\sim10^{14}M_{\odot}$ virialised cluster by $z=0$. As progenitors of present-day clusters hosting massive ellipticals (which almost always host radio galaxies; \citealt{saba19}), they give information on the evolution of radio galaxies, their plasma, and their hosts over cosmic time. Understanding of the radio plasma and the interaction with their environments gives information on feedback at a relatively unexplored epoch.

An important probe of this plasma comes from X-rays, which in usual cases describe scattering of background photons by electrons in the lobe through the inverse-Compton (IC) process. Several radio-X-ray studies of low-redshift radio galaxies, revolutionised by the advent of \textit{Chandra}, have concluded that the background photon field being up-scattered is either the synchrotron photons from the radio plasma itself (synchrotron self-Compton; SSC), or the ever-present CMB photons (ICCMB). The latter of the two is most common in the low-energy diffuse regions of the lobes \citep{cros05,hard07,gill21}. In either case, one gains synergy in information on the physical conditions inside the lobes with both radio and X-ray measurements of the same particle populations. 

While the X-rays could also describe synchrotron emission from high-energy particle acceleration \citep[e.g.][]{hard07}, it's clear on energetic grounds that they do not dominate the energetics overall on a population level \citep{gill21}. Meanwhile, the nature of the low-energy electron population driving ICCMB X-rays has remained elusive. Since radiative losses preferentially deplete the highest-energy lobe electrons of energy, the radio emission (at GHz frequencies) in the lobes fades more rapidly than the X-rays that describe ICCMB from lower energy lobe electrons. This leads to the appearance of so-called `IC ghosts' with extended X-rays but no detectable radio emission associated with AGN lobes \citep{mocz11}. Their characteristic composition of low-energy electrons means that they are important in tracing the longevity of the radio galaxy plasma, playing a role in AGN feedback, as well as in producing the Sunyaev–Zeldovich (SZ) effect in a cosmological context \citep{cola08}. Crucially, IC Ghosts are clearly unaccounted for in high flux-limited radio surveys, and therefore in studies using radio galaxy populations drawn from them. 
\begin{figure*}
 \label{fig:3c294}
    \centering
    \hspace{-0.6cm}
    \includegraphics[width=0.5\linewidth]{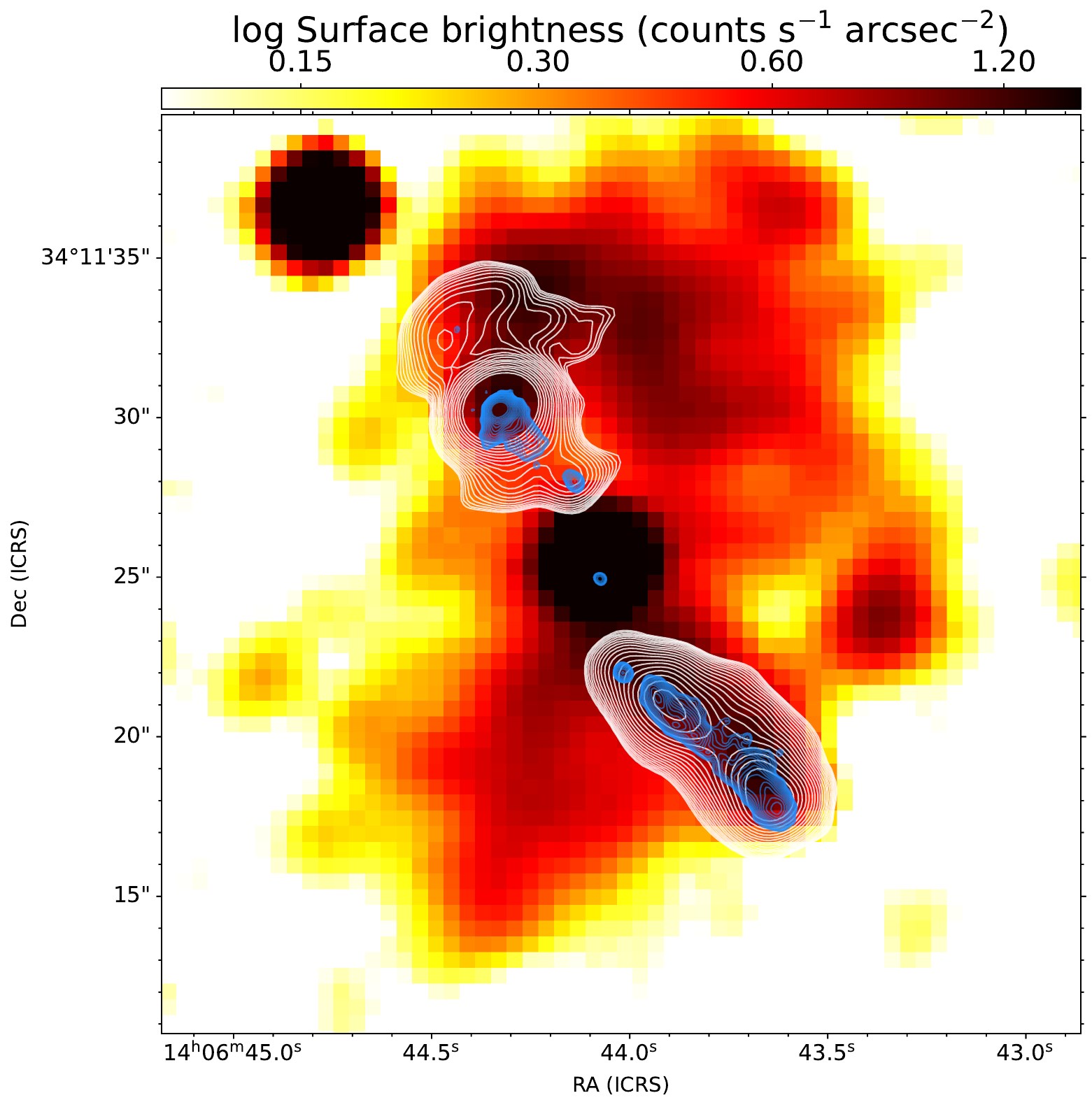}\hspace{-0.1cm}
    \includegraphics[width=0.5\linewidth]{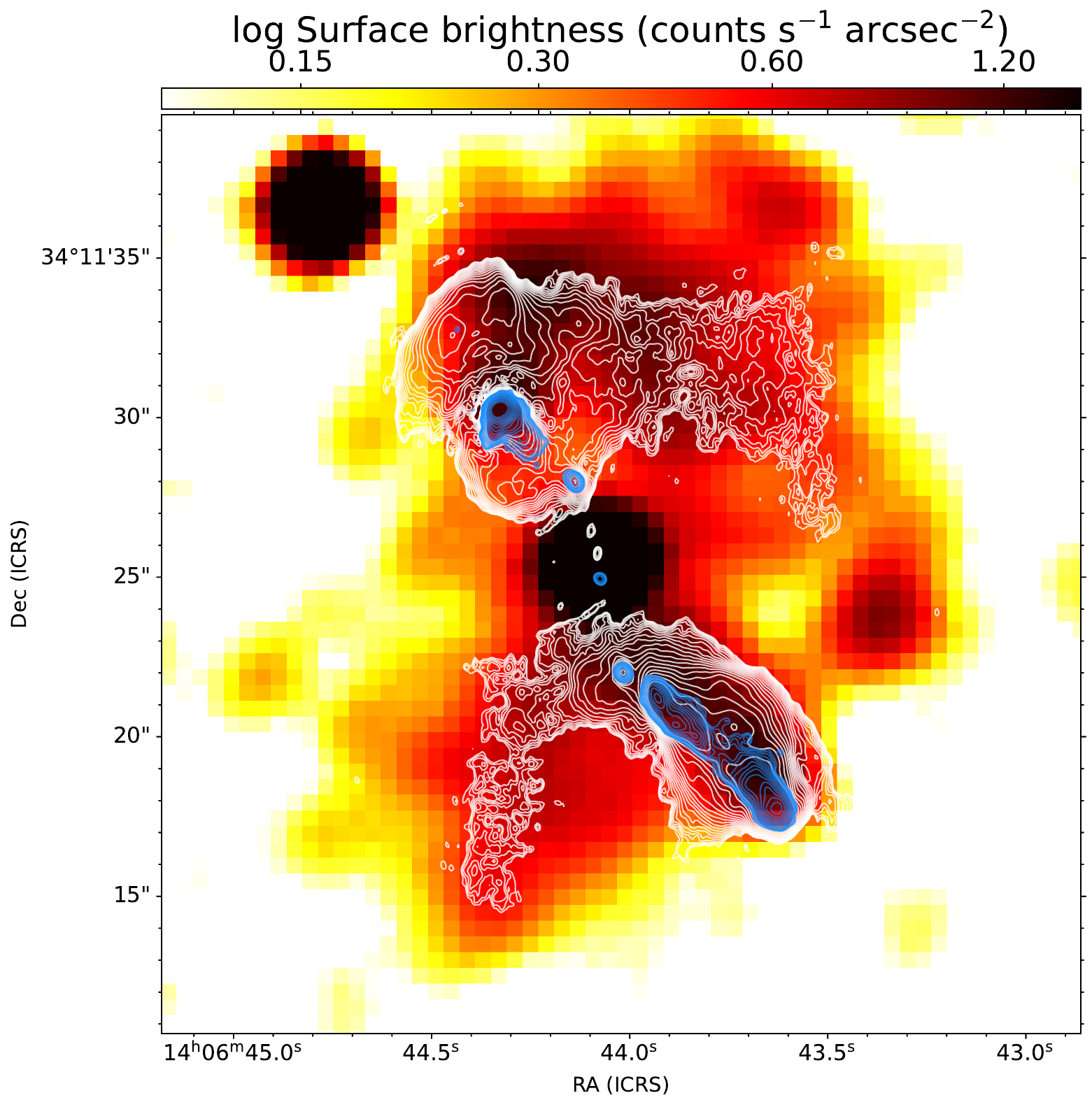}
    \caption{\textit{Chandra} X-ray image of 3C\,294, adapted from \citealt{fabi03} and \citealt{erlu06}, overlaid by radio contours at different frequencies. Left: VLA 1.4\,GHz (white) and VLA 5\,GHz contours (blue). Right: LOFAR 144\,MHz (white) and VLA 5\,GHz contours (blue). Contours are 3$\sigma$ multiplied by 2$^n$ where $n=[0,1,2...10]$, except for the blue contours, which start at 80$\sigma$. Radio contours have an angular resolution $\sim0.3$\,arcsec, except the VLA 1.4\,GHz (left) at 1.2\,arcsec. The reprocessed 5\,GHz data is sensitive to the radio core, unseen in previous publications. The low-energy plasma at 144\,MHz (right) coincident with the diffuse X-ray emission is clearly detected, along with knots along the northern jet between the core and the northern inner hotspot.}
   
\end{figure*}

Since the CMB temperature scales as $(1+z)^4$, cancelling the effect of dimming due to distance, one expects an abundance of IC ghosts in sensitive X-ray surveys. Identifications of IC Ghosts have, however, remained elusive: snapshot surveys of high redshift radio-loud objects \citep[e.g.][]{stua18} have failed to detect more than a handful of extended X-ray emission (associated with extended lobe radio emission\footnote{\cite{jime20} detect 6 out of 9 radio-loud objects to have X-ray emission in the lobes or hotspots, but without sufficient angular resolution to distinguish between the two.}) over the compact X-ray emission from the AGN cores and hotspots. Meanwhile, simulations by \citealt{mocz11} predict an over-abundance of IC Ghosts relative to what is observed: up to 30\% of double-lobed extended X-ray emission is predicted to come from IC Ghosts. This prohibits important information on radio galaxies in general: without co-spatial radio and X-ray detections (or observations sensitive to the emission, at least) of the lobes, we cannot make statements about the number density of radio galaxies as a whole as a function of redshift. \citealt{mocz11} also predict radio lobes from powerful jets occupy up to 30\% of the volume filling fraction of cosmic filaments during the quasar era ($z\sim2$). However, these predictions are sensitive to the assumptions on the lobe energetics (e.g. the distribution and range of the electron Lorentz factors, and the magnetic field strength in the plasma). Measurements of both synchrotron and inverse-Compton emission from the same particle population allow direct measurements of the magnetic field strength, alleviating degeneracy and allowing accurate calculations of the energy in the lobes \citep[e.g.][]{harr79,hard98,cros05}. Otherwise, an equipartition assumption is made on the magnetic field energy density contributing to the synchrotron emission we observe, which is known to be inaccurate  \citep{cros18,maha20}.

Constraining the low ($\gamma_{\text{min}}$) and high energy ($\gamma_{\text{max}}$) limits to the electron energy distribution that defines the integrated synchrotron spectrum of lobes remains challenging. Radio detections of IC Ghost plasma can place useful constraints on these limits, which in turn influence the inferred energy content in the lobes. The energy density of an electronic population in the lobes is given by
\begin{equation} 
 E=\int_{\gamma_{\text{min}}}^{\gamma_{\text{max}}}d\gamma\,n_{\text{e,rel}}(\gamma)\gamma m_{e}c^2
\end{equation}
where the power-law electron energy distribution $n_{\text{e,rel}}(\gamma) =n_{\text{e,0}}\gamma^{-p}$. For typically steep energy spectra in the lobes ($p>2$), the total energy is most sensitive to $\gamma_{\text{min}}$.
The absence of observational constraints requires an assumption for the value of $\gamma_{\text{min}}$, which is also sensitive to the prediction of the presence of IC Ghosts in models (\citealt{mocz11}). \cite{erlu06} and \cite{fabi09} argue for a low-energy cut-off $\gamma_{\text{min}}\sim10^3$ in the lobes of the IC Ghost HDF\,130, far higher than competing arguments which suggest a range $\sim10^0-10^2$ \citep[e.g.][]{hard05_microq,kais97_pd}. Lower values are energetically favourable, since in hotspots $\gamma_{\text{min}}\sim10^2-10^3$ can be inferred \citep[e.g.][]{hard01}. After acceleration at the hotspots, lobe electrons undergo adiabatic and radiative losses that shift the particle distribution toward lower $\gamma$, though fundamentally $\gamma_{\text{min}}$ is determined by the particle acceleration conditions \citep{blan87}. Without observational measurements of $\gamma_{\text{min}}$ and the energy index $p$ (which can be made by fitting $n_{\text{e,rel}}(\gamma)$ to broad-band radio measurements), calculations of the total lobe energy are unreliable. 

Sensitive radio and X-ray surveys over a wide-sky area give us the opportunity to address these important questions. However, at low redshifts, the CMB energy density is low, and other photon fields, such as that from the central AGN, may dominate IC emission at X-ray energies. Lorentz factors of $\gamma\sim10^3$ are required to up-scatter the CMB photons to X-ray energies that we perceive at $\sim$1\,keV at $z=0$, and these electrons tend to radiate synchrotron at $\lesssim$100\,MHz. Due to the lack of sensitive low-frequency telescopes in the past, studies of inverse-Compton scattering have been limited to higher ($\gtrsim$1\,GHz) frequencies \citep[e.g.][]{brun97,lask10}, which do not have the sensitivity to this emission (a problem which is exacerbated at higher redshifts). This has resulted in a situation where the detections of probable IC Ghosts from current X-ray facilities (e.g. 3C\,432, 3C\,294 and 3C\,191; \citealt{erlu06}) lack detections of the spatially-coincident radio emission from synchrotron light, particularly in a resolved manner that allows one to distinguish between the lobe material and the high energy jet and hotspot-related material.  

3C\,294 ($z=1.786$) in particular is an interesting case, as the GHz-frequency emission shows two pairs of hotspots either side of a core, suggesting restarting activity, but the outer hotspots subtend at an angle of 20 degrees on the plane of the sky relative to the inner hotspots, suggesting jet precession is also occurring, as first suggested by \cite{mcca90}. A dense Lyman-$\alpha$ environment ($L_{\alpha}\sim8\times10^{44}$\,erg s$^{-1}$; \citealt{mcca90}) combined with evidence of clumpy optical structure and several optical counterparts indicative of a merging or dual-AGN system\footnote{This has come into question with new spectroscopic data which suggest a background AGN instead \citep{quir21}.} \citep{quir01}, suggest the source is embedded in a dense, dynamic protocluster environment. A 20\,ksec X-ray map with \textit{Chandra} revealed extended and diffuse hourglass morphology extending 100\,kpc to the north and south of the central core \citep{fabi01}, suggesting a central ICM emitting the X-rays through Bremsstrahlung. Analysis of deeper X-ray data suggested an ICCMB origin \citep{fabi03} using an absorbed power-law model with a photon index of $\Gamma\sim2.3$. The lack of radio emission associated with the extended X-ray emission, even with deep 1.4\,GHz data \citep{erlu06}, suggests the presence of weak magnetic fields and/or very low Lorentz factors, if there is co-spatial synchrotron plasma at all.

In this paper, we present a study using new low-frequency ($144$\,MHz) observations of 3C\,294. For the first time, we detect the low-energy lobe emission in the radio (Figure \ref{fig:3c294}), associated with the extended X-rays, allowing constraints on the plasma physical conditions. We model the synchrotron and ICCMB emission using the available radio and X-ray data, and determine the physical conditions of the plasma that lead to the observable emission. While we present an analysis of the lobe material responsible for the diffuse X-ray emission, a robust multi-wavelength analysis of the jet components and hotspots will be presented in a forthcoming paper (Mahatma+, in prep). We use a flat $\Lambda$CDM cosmology where $H_0$=69.6\,km\,s$^{-1}$\,Mpc$^{-1}$ and $\Omega_{M}=0.286$ and $\Omega_{\text{vac}}=0.714$.

\section{Data}
\label{sect:data}
\subsection{Radio}
We utilise observations of 3C\,294 made as part of the Low Frequency Array (LOFAR; \citealt{vanh13}) Two-Metre Sky Survey (LoTSS; \citealt{shim22}), which comprises 8-hour pointings observed at 120-187\,MHz, using the full pan-European array (giving angular resolutions down to 0.2\,arcsec). 3C\,294 was observed in pointing P210+35 (observation ID L708094, project ID LT10\_010), along with the primary flux calibrator 3C\,196. The data were processed using the LOFAR Initial Calibration (LINC) pipeline\footnote{\url{https://git.astron.nl/RD/LINC}} \citep{dega19} and the DDF pipeline\footnote{\url{https://github.com/mhardcastle/ddf-pipeline}} \citep{shim19,tass21}, which mainly determine and apply calibration solutions for the Dutch stations of LOFAR. To calibrate the `international' stations outside the Netherlands, we use the \texttt{LOFAR-VLBI} pipeline \citep{mora22} which makes use of the \texttt{facetselfcal}\footnote{\url{https://github.com/rvweeren/lofar_facet_selfcal/}} strategy -- this performs self-calibration of the international stations using the core Dutch stations phased-up to provide a single, sensitive station as reference. For an initial skymodel for self-calibration, we use an 8\,GHz model image (see below) to constrain the positions of the most compact components. Several rounds of self-calibration led to a sensitive low-frequency map (using \texttt{WSCLEAN}) centred at 144\,MHz, with an angular resolution of $\sim0.3$\,arcsec and a background RMS noise of 70\,$\mu$Jy beam$^{-1}$. The dynamic range of the map, presented in Figure \ref{fig:3c294} (white contours, right panel), is $\sim15400$.

For our analysis, we utilise archival higher frequency radio data at 1.4\,GHz, 3\,GHz, 5\,GHz, 8.4\,GHz and 15\,GHz, all available in the NRAO archive. We process each dataset in \texttt{CASA} in the standard manner, with several rounds of self-calibration. The 15\,GHz A-array and B-array data were combined in the $uv-$plane before imaging. We summarise the observational data used in this work in Table \ref{tab:obs}. We note that the data at 5\,GHz and above are only sensitive to angular scales $\lesssim10$\,arcsec, smaller than the angular extent of 3C\,294 ($\sim20$\,arcsec), and are thus not sensitive to the diffuse low surface brightness plasma shown by the LOFAR image in Figure \ref{fig:3c294}. These data are only used in our analysis of the jets and hotspots in Section \ref{sect:hotspots}.

We overlay the 1.5\,GHz and 5\,GHz VLA maps (white and blue contours in the left panel, respectively) and the 144\,MHz LOFAR map (white contours; right panel) over the \textit{Chandra} 0.3-6.0\,keV X-ray map presented by \citep{fabi03} in Figure \ref{fig:3c294}. The previously unseen diffuse lobe emission west of the northern hotspot and east of the southern hotspot, coincident with the extended X-ray emission, is clearly detected. Through our re-processing, we also detect the (previously unseen) core at 5\,GHz, while the LOFAR data also shows faint jet knots between the 5 GHz core and the northern inner hotspot, describing the current jet. We make the distinction between hotspots and knots here based on morphological grounds: jet knots are footprints of a continuing jet outflow, and are susceptible to relativistic effects (i.e. Doppler boosting), making them appear only on one side of the radio source. Hotspots, on the other hand, are suggested to describe non-relativistic shocks \citep{blan78}, and are therefore more commonly seen symmetrically placed either side of a radio core as seen in Figure \ref{fig:3c294}. There are objects where these descriptions do not hold \citep[see][and references therein]{hard08}, and the current data do not rule out that the inner hotspots are actually knots from the same jet flow.
\begin{table}
    \centering
    \begin{tabular}{c|c|c|c|c}
        Frequency & Telescope & LAS & Project code \\ \hline
        144\,MHz & LOFAR international array & 70\,arcsec$^{\dagger}$ & LT10\_010\\
        1.5\,GHz & VLA A-array & 36\,arcsec & AP158\\
        3\,GHz & VLA BnA-array & 58\,arcsec & 20B-375\\
        5\,GHz & VLA A-array & 9\,arcsec & AM224\\
        8.4\,GHz & VLA A-array & 6.5\,arcsec & AB917\\
        15\,GHz & VLA A-array & 3.6\,arcsec & 20B-356\\
        15\,GHz & VLA B-array & 12\,arcsec & 21B-356 \\ \hline

    \end{tabular}
    \caption{Radio data processed and used in this study. The two 15\,GHz datasets were jointly-deconvolved in the $uv-$plane to match the baseline coverage of other datasets when imaging at sub-arcsecond resolution. The LAS column refers to the largest angular scale of emission detectable with that setup of observations. $\dagger$ The LOFAR inner Core stations were combined before self-calibration, as mentioned in the text, providing a smaller effective angular scale than possible without station combination -- this does not affect the sensitivity to large-scale components in 3C\,294, which itself is only 20\,arcsec in size.}
    \label{tab:obs}
\end{table}
\subsection{X-ray}
We re-process the archival \textit{Chandra} data of 3C\,294 (observation IDs 3207 and 3445) used by \cite{fabi03}, and we follow their processing strategy. The two sets of observations were taken on 25th February 2002 with an exposure of 69.8\,ks, and the second on 27th February 2002 with an exposure of 122.0\,ks, with a combined effective exposure of 191.8\,ks. To process the data into level 2 events files, we use the latest calibration tables using the \texttt{chandra\_repro} script as part of \texttt{CIAO v4.17}, individually for both data sets. We checked the astrometry of both data sets using the \texttt{fine\_astro} task, and the resulting images were spatially consistent (well within the \textit{Chandra} PSF). To extract the spectra in the diffuse regions, we use the same rectangular apertures (length $\sim$15 arcsec, radius $\sim$5 arcsec) as used by \cite{fabi03} and displayed in Figure \ref{fig:3c294_lowres}, which maximises the number of counts extracted. The positions of the boxes were chosen to capture the diffuse X-ray emission while evading the compact regions (i.e. the jet and hotspots) that are also thought to produce IC emission. We also tested ellipsoidal regions, which only encompass the extent of the diffuse emission at 144\,MHz, but these smaller regions resulted in too few X-ray counts per spectral bin to allow statistically significant fits of physical models to the spectra. For a background region, we use a 30\,arcsec-radius circle in a blank region of the CCD -- we alternatively tested an annulus around 3C\,294 (as used to subtract foreground thermal material of rich-cluster objects by \citealt{maha20}), but this resulted in too many counts subtracted for robust spectral fitting. We then extracted the spectra of the diffuse regions using \texttt{specextract}, using the option to generate weighted source and background response files (since we measure extended emission), and then combine the spectra for the two datasets using \texttt{combine\_spectra}, binning the data into spectral bins containing at least 20 counts as performed by \cite{fabi03}. Note, we also attempted to extract spectra of the northern diffuse region solely, but this also resulted in insufficient counts. 

\begin{figure}
    \centering
    \includegraphics[width=\linewidth, trim={0cm 0.2cm 0.8cm 1.36cm}, clip]{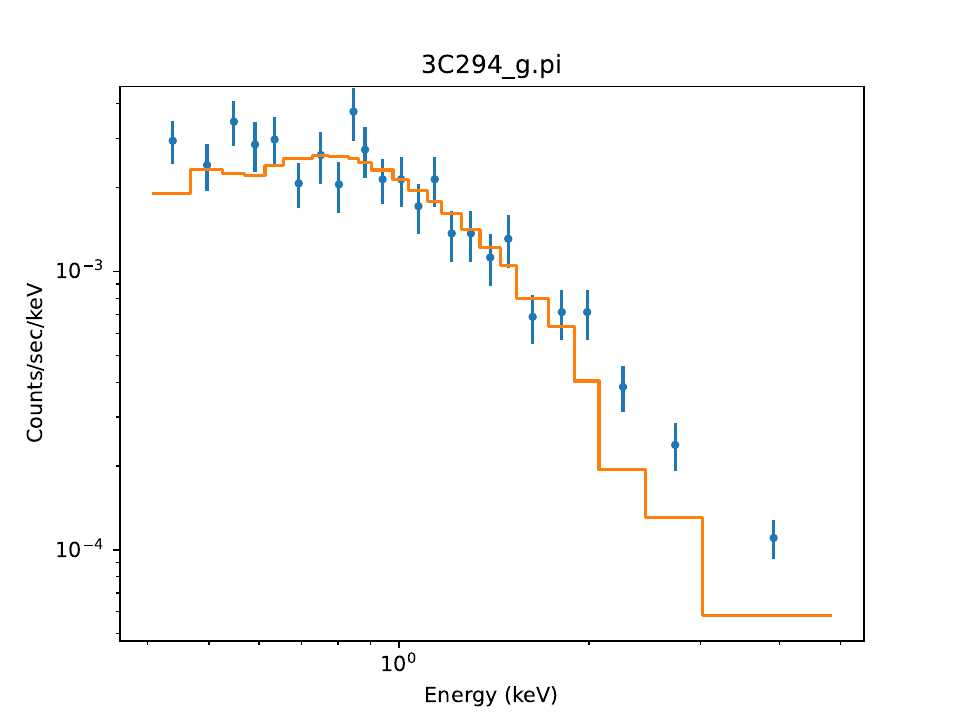}
    \caption{X-ray 0.3-6\,keV spectrum of the diffuse regions shown in Figure \ref{fig:3c294_lowres}, fitted with an absorbed (\texttt{PHABS*ACISABS}) non-thermal power-law model (orange line).}
    \label{fig:sherpa_fit}
\end{figure}

We used the \texttt{sherpa} application to construct and fit emissivity models to the extracted spectra of the diffuse regions. After ignoring bad pixels and limiting the energy range to 0.3-6.0\,keV, we fit the spectra with an XSPEC power-law model (\texttt{xspowerlaw}; to reflect the inverse-Compton emission) with a \texttt{PHABS} absorbing screen, keeping the three parameters (line of sight neutral hydrogen, power-law index, and power-law normalisation) free. We used \texttt{w-stat} for the fitting, which includes the background. The results of the fit, with a reduced $\chi^{2}_{\text{red}}=0.94$, are tabulated in the top row of Table \ref{tab:sherpa_fit}. The power-law values are broadly consistent (we find a fitted photon index of $\Gamma=2.10\pm0.23$ versus the \citealt{fabi03} result of $\Gamma=2.3\pm0.3$), however the value for the hydrogen column density $N_H$ is poorly constrained -- we follow the strategy of \cite{fabi03} by further absorbing the power-law with the \texttt{ACISABS} model (accounting for the known degradation of the ACIS soft X-ray response to correct for instrumental effects), and freezing the \textit{PHABS} hydrogen column density to the Galactic value ($N_H=1.2\times10^{20}$\,atoms\,cm$^{-2}$). This provides smaller errors on the power-law parameters (with a photon power-law index of $\Gamma=2.59\pm0.10$ and a 1\,keV flux density of $3.08\pm0.14$\,nJy; middle row of Table \ref{tab:sherpa_fit}) with $\chi^{2}_{\text{red}}=1.20$. Although an absorbed power-law without the \texttt{ACISABS} model (but with $N_H$ fixed to the Galactic value) provides a better fit ($\chi^{2}_{\text{red}}=0.8$), the well-documented degradation of the ACIS and the associated instrumental effects were mitigated only after the observations were taken \citep{gran24}, and so we use the fits with the \texttt{ACISABS} model included. The fitted spectrum is shown in Figure \ref{fig:sherpa_fit}. We use this inverse-Compton flux density in our modelling of the broad-band emission of the diffuse region in Section \ref{sect:analysis}. Note that we also tested an XSPEC \texttt{mekal} model to fit thermal bremsstrahlung models, but this resulted in a poorer fit in the bottom row of Table \ref{tab:sherpa_fit}, while a thermal gas model for the diffuse emission is strongly disfavoured on morphological grounds \citep[see][]{fabi03}.
\begin{table}
    \centering
    \begin{tabular}{c|c|c|c}
    Parameter & Fitted value & Unit & $\chi^{2}_{\text{red}}$\\ \hline
    \texttt{PHABS}*\texttt{powerlaw}& & & 0.94\\ \hline
    $N_H$ & $(2.70\pm42.41)\times10^{19}$ & atom\,cm$^{-2}$  \\
    $\Gamma$ & $2.10\pm0.23$ & \\
    Power-law 1\,keV flux & $2.30\pm0.36$ & nJy \\ \hline
    \texttt{PHABS}*\texttt{ACISABS}*\texttt{powerlaw}$^{\dagger}$ & & & 1.20 \\ \hline
    $\Gamma$ & $2.59\pm0.10$ & \\
    Power-law 1\,keV flux & $3.08\pm0.14$ & nJy \\ \hline
    \texttt{PHABS}*\texttt{ACISABS}*\texttt{MEKAL}$^{\dagger}$ & & & 2.04 \\ \hline
    $kT$ & $2.61\pm0.20$ & keV \\
    Thermal abundance & $1.66\pm0.98$ & $Z_{\odot}$ \\
    Thermal 1\,keV normalisation & $(9.20\pm2.51)\times10^{-5}$ & \\ \hline
    \end{tabular}
    \caption{Table of free parameters and their fitted values using the absorbed power-law models fitted to the X-ray spectra of the diffuse regions. $\dagger$ $N_H$ is fixed to the Galactic value of $1.2\times10^{20}$\,atom\,cm$^{-2}$. All measurements are given with a 1$\sigma$ confidence level.}
    \label{tab:sherpa_fit}
\end{table}

\section{Analysis}
\label{sect:analysis}
\subsection{Modelling the diffuse plasma}
\label{sect:modelling}
\begin{figure}
    \centering
    \includegraphics[width=1\linewidth]{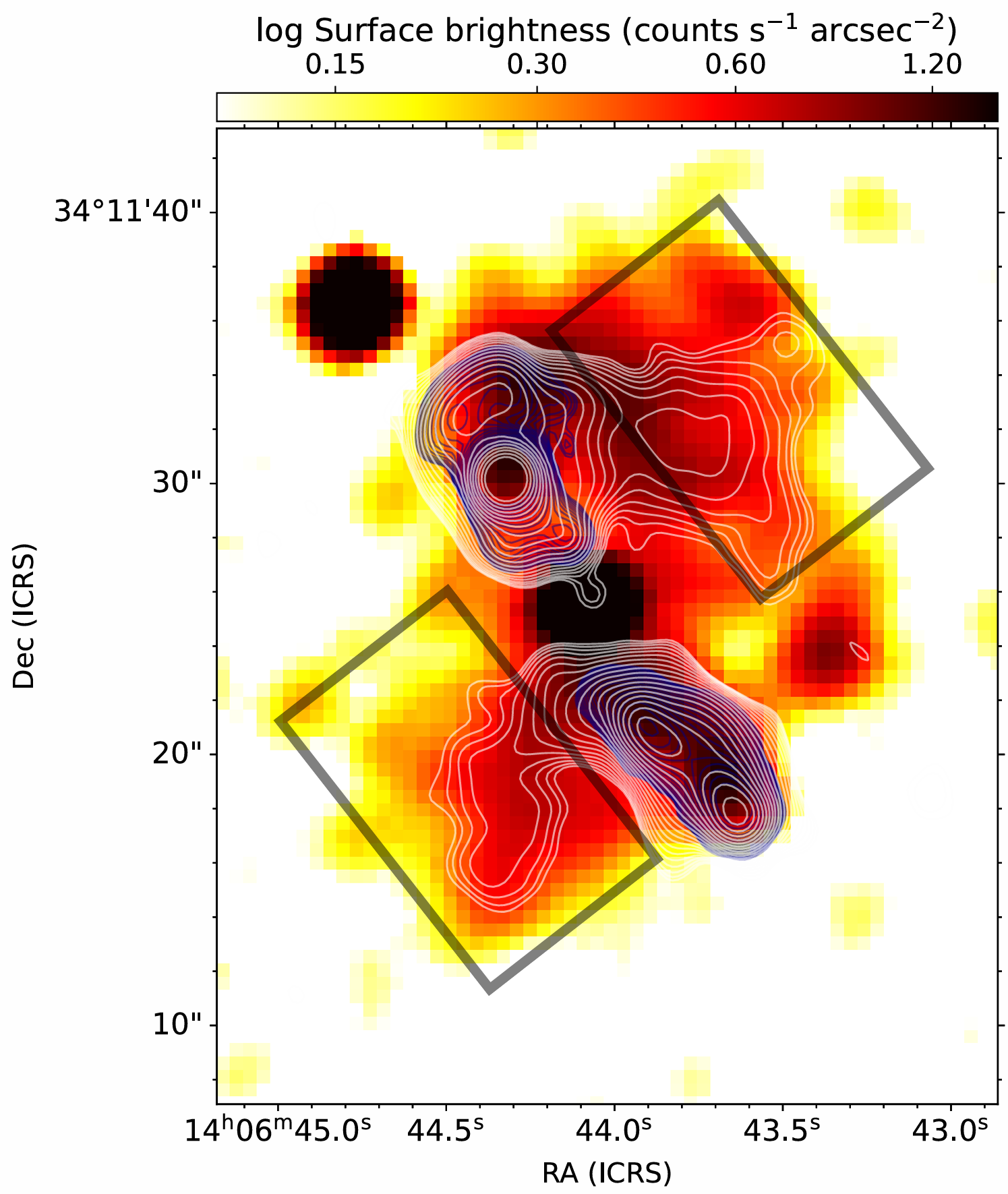}
    \caption{Same as in Figure \ref{fig:3c294} but with the LOFAR image convolved at 1.2\,arcsec resolution (white contours) using $uv-$tapering and with the 1.5\,GHz VLA data (blue contours) convolved to the same resolution. The grey rectangles describe the regions used for the X-ray spectral fits by \citet{fabi03}, which we also use for flux measurements of the 144\,MHz emission.}
    \label{fig:3c294_lowres}
\end{figure}

To model the synchrotron and inverse-Compton spectra in the diffuse part of the radio and X-ray lobe emission, we make radio flux measurements in the rectangular apertures as used to fit a non-thermal model to the X-ray spectra (grey dashed boxes in Figure \ref{fig:3c294_lowres}) described in Section \ref{sect:data}. We measure the flux density in the same regions at 144\,MHz (convolved to a 1.2\,arcsec resolution to match the limits we also obtain at 1.4\,GHz and 3\,GHz, measured with the same apertures). The radio and X-ray measurements are tabulated in Table \ref{tab:fluxes}, as well as upper limits at high radio frequencies for completeness.

%Using the LOFAR flux density in this region, we calculate an equipartition magnetic field of 6.6$\times10^{-9}$\,T (6.6\,nT or 66\,$\mu$G), using a single power-law model, a minimum Lorentz factor $\gamma_{\text{min}}=1$, a maximum Lorentz factor $\gamma_{\text{max}}=10^8$, and an electron energy index of $p=3.6$ (equivalent to the measured photon index of $\Gamma=2.3$ ;\citealt{fabi03}). We note that this is \textit{generally} an upper limit, as is usually the case for equipartition in radio galaxy lobes \citep{cros18}. The predicted inverse-Compton flux at 1\,keV for this model is 0.07\,nJy, nearly two orders of magnitude below our \textit{Chandra} 1\,keV measurement (Table \ref{tab:fluxes}). This simplistic model assumes a power-law electron energy distribution at all radio frequencies -- we already know that the lobes have a steep spectrum and there must be a spectral break at GHz (or lower) frequencies. If we assume the break in the synchrotron spectrum occurs at 1\,GHz (at and beyond which current radio maps are not sensitive to detect the emission), then using the critical frequency formula for synchrotron radiation, $\nu_c = \frac{3}{4\pi} \,\gamma^2 \,\frac{e B}{m_e c}$, the Lorentz factor at which the electron energy distribution breaks is $\gamma\sim8\times10^{3}$, and therefore a single power-law up to $\gamma\sim10^8$ does not apply. 

\begin{figure}
    \centering
    \includegraphics[scale=0.57, trim={0cm 0.2cm 0cm 1cm},clip]{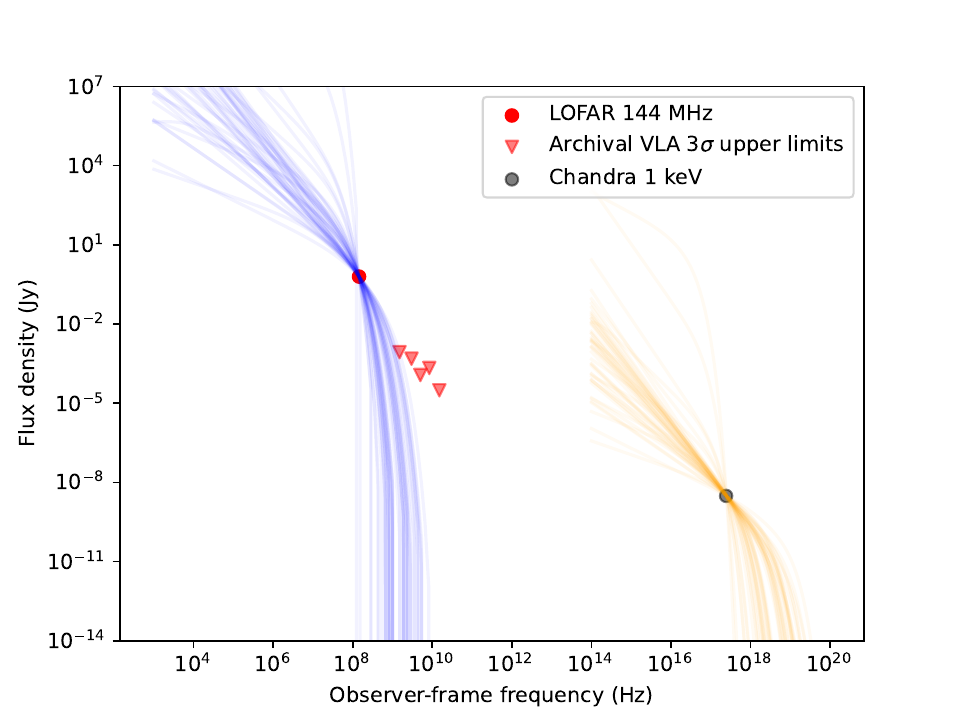}
    \caption{Modelled spectra using the parameters in Table \ref{tab:model_params}, fitted to the measurements in Table \ref{tab:fluxes}. Blue curves show the synchrotron emission, and yellow curves show the IC emission, for models with physical parameters consistent with our observational constraints. Red arrows represent upper limits at GHz frequencies, based on 3$\sigma$ of the box regions presented in Figure \ref{fig:3c294_lowres}. The upper limits at 5\,GHz and above are for reference only, and not used to constrain the spectra.}
    \label{fig:plot_spec}
\end{figure}
To quantify the probable distribution of physical parameters based on our measurements, we model the plasma as a broken power-law by fitting models for synchrotron and inverse-Compton for a range of parameters (using the \texttt{PYSYNCH}\footnote{https://github.com/mhardcastle/pysynch/} code) to our observational constraints (144\,MHz, 0.3-6.0\,keV, and radio upper limits at 1.4\,GHz and 3\,GHz; Table \ref{tab:fluxes}). To determine the emission, we draw values from log-uniform distributions of the following unknown parameters: $\gamma_{\text{min}}$, $\gamma_{\text{max}}$, $\gamma_{\text{break}}$, $\Delta p$ (i.e. the change in power-law index after the break), and linear-uniform distributions of the magnetic field strength $B$, and the injection index $p$. Table \ref{tab:model_params} shows the ranges for their values. The only fixed physical parameters are the redshift $z=1.786$, and the volume of the emitting region (set to two cylinders of length 15.3\,arcsec and radius 5.2\,arcsec, as used to measure the radio and X-ray emission). The volume filling factor is set to unity, as a default in the code. While this is a common assumption, if the true filling factor $\phi$ is less than 1, then the predicted electron density in the lobe increases as $n_e\propto\phi^{-\frac{1}{2}}$, as a caveat to our results. We randomly draw values from the unknown parameters (discarding parameter sets where $\gamma_{\text{min}}\geqslant\gamma_{\text{break}}\geqslant\gamma_{\text{max}}$, to neglect unphysical models) and use \texttt{PYSYNCH} to calculate the synchrotron and IC emissivities, normalizing the spectrum in each case to the 144\,MHz measurement. We do not account for Doppler boosting, which is likely not occurring for the aged non-relativistic plasma in question. We produce 100,000 emissivity models to ensure stochasticity.

\begin{table}
    \centering
    \begin{tabular}{c|c}
    Data & Flux density (Jy)\\ \hline
    LOFAR 144\,MHz & 0.6255$\pm0.0005$\\%0.1662+0.4953 from SE and NW lobe 0.426+0.153
    VLA 1500\,MHz & $<8.541\times10^{-4}$\\%rms of 2.064e-4 and 1.961e-4
    VLA 3000\,MHz & $<4.846\times10^{-4}$ \\ %rms of 1.018e-4 and 1.254e-4
    VLA 5000\,MHz & $<1.167\times10^{-4}$ \\ %rms of 1.018e-4 and 1.254e-4
    VLA 8400\,MHz & $<2.149\times10^{-4}$ \\ %rms of 1.018e-4 and 1.254e-4
    VLA 15000\,MHz & $<3.082\times10^{-5}$ \\ %rms of 1.018e-4 and 1.254e-4
    \textit{Chandra } 1\,keV & $3.081\times10^{-9}\pm0.143$\\
    \end{tabular}
    \caption{Observer-frame flux density measurements and upper limits of the diffuse plasma measured in the rectangular regions shown in Figure \ref{fig:3c294_lowres}. Upper limits are based on the quadrature sum of the 3$\sigma$ RMS values in those regions. Errors represent statistical errors -- background RMS noise and spectral fitting for the LOFAR and \textit{Chandra data}, respectively.}
    \label{tab:fluxes}
\end{table}
\begin{table}
    \centering
    \begin{tabular}{c|c|c}
    Parameter & Range & Median fitted\\ \hline
    $\gamma_{\text{min}}$ & 1 - $10^3$ & 1.13\\
    $\gamma_{\text{max}}$ & 1 - $10^8$ & 1447.12\\
    $\gamma_{\text{break}}$ & 1 - $10^8$ & 2.98\\
    $B$ (T) & $1\times10^{-10} - 1\times10^{-7}$ & $2.80\times10^{-9}$\\
    $p$ & 2 - 4 & 3.09\\
    $\Delta p$ & 0.5 - 2 & 1.36\\
    \end{tabular}
    \caption{Model parameters drawn from uniform distributions in the stated ranges. The median fitted column states the medians of the distributions of parameters that produce emission consistent with our observational data for 3C\,294.}
    \label{tab:model_params}
\end{table}
\begin{figure*}
    \centering
    \includegraphics[width=0.5\linewidth]{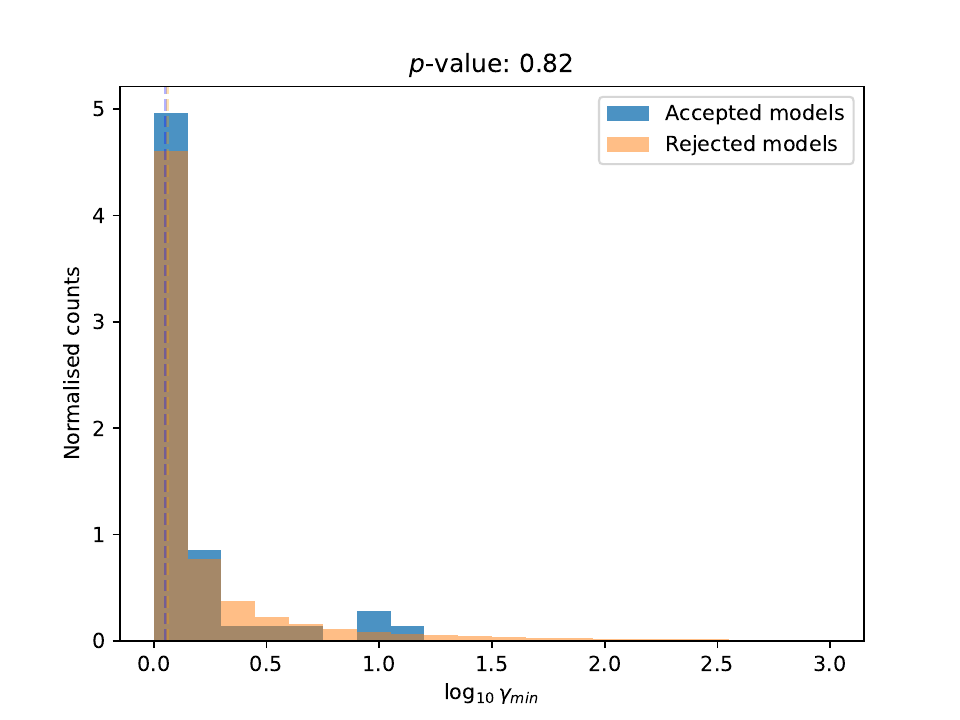}\hspace{-0.5cm}
    \includegraphics[width=0.5\linewidth]{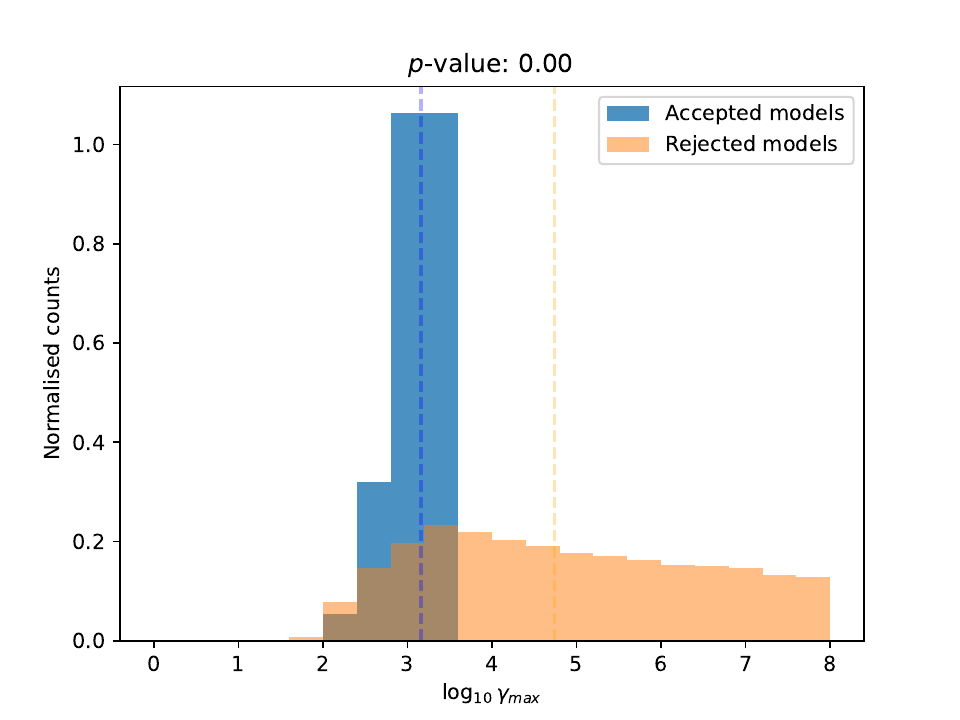}
    \includegraphics[width=0.5\linewidth]{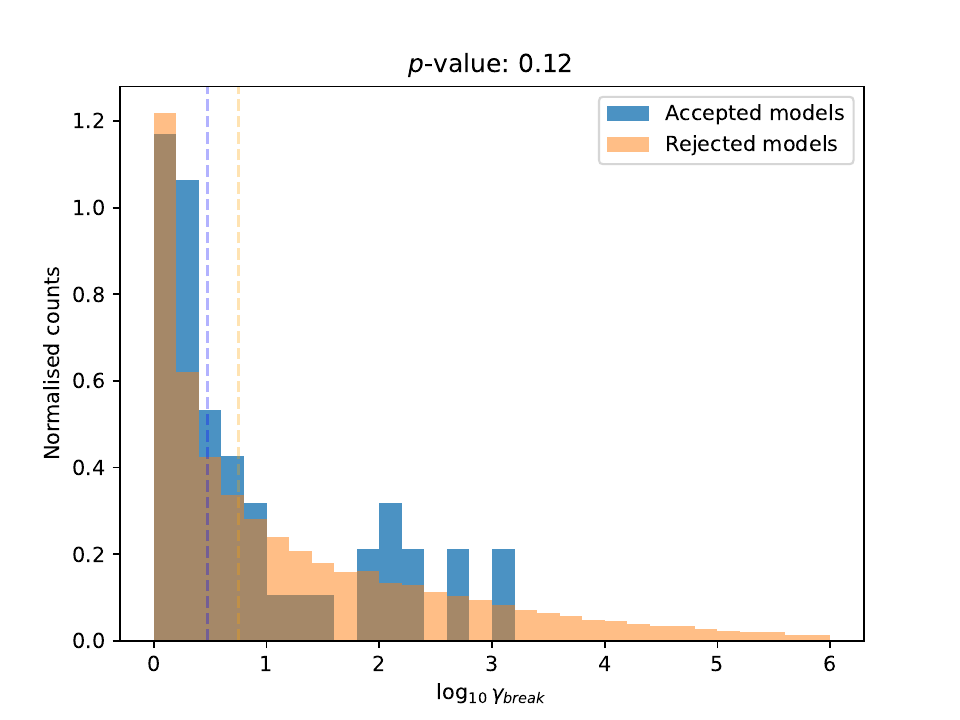}\hspace{-0.5cm}
    \includegraphics[width=0.5\linewidth]{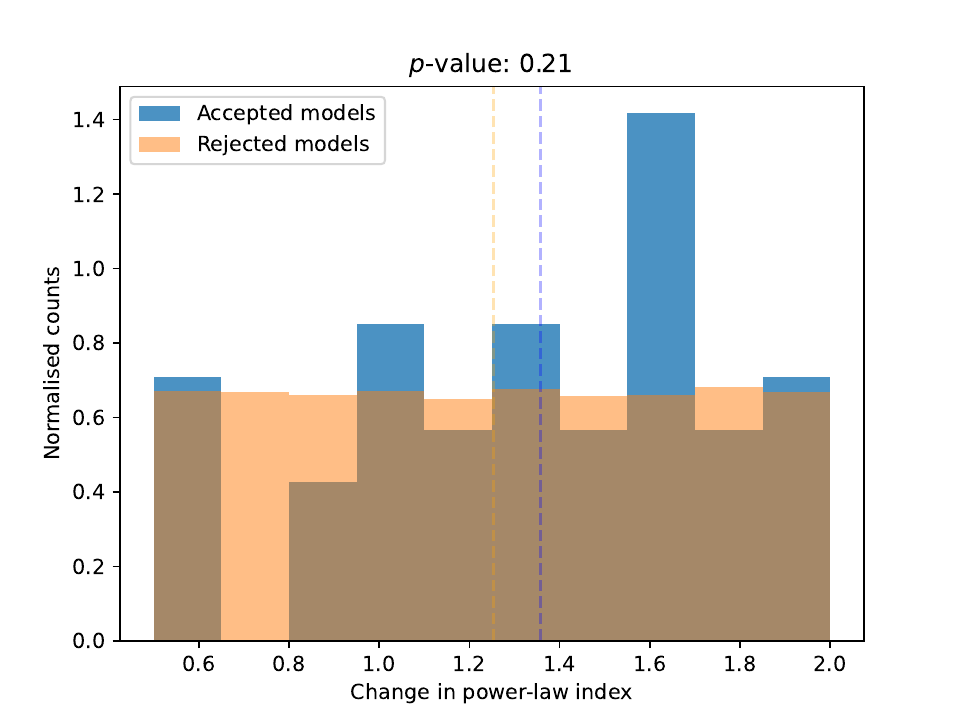}
    \includegraphics[width=0.5\linewidth]{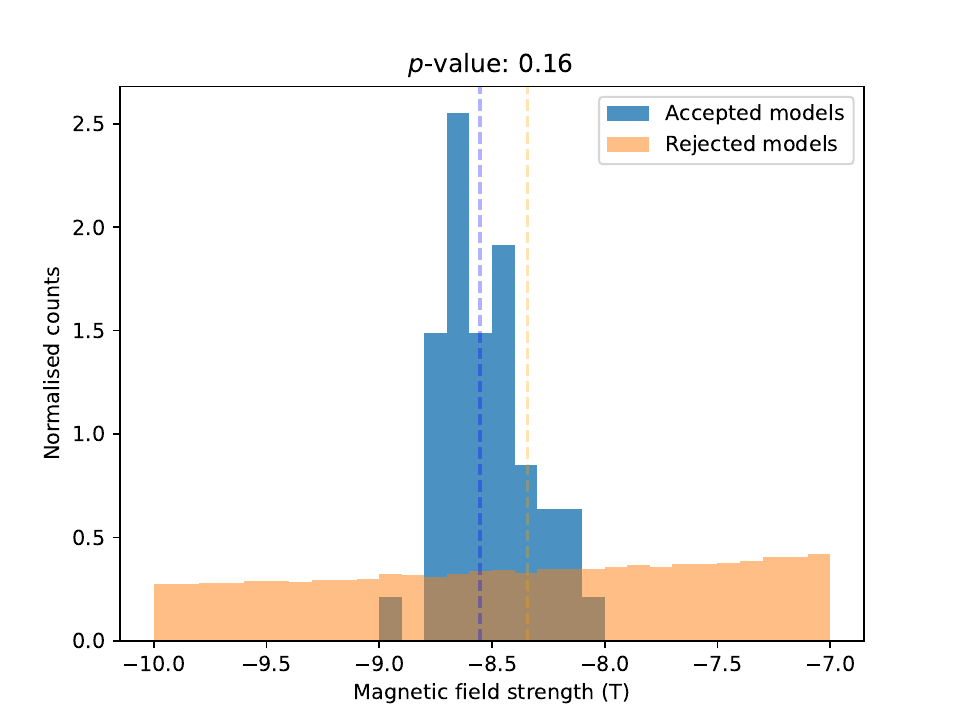}\hspace{-0.5cm}
    \includegraphics[width=0.5\linewidth]{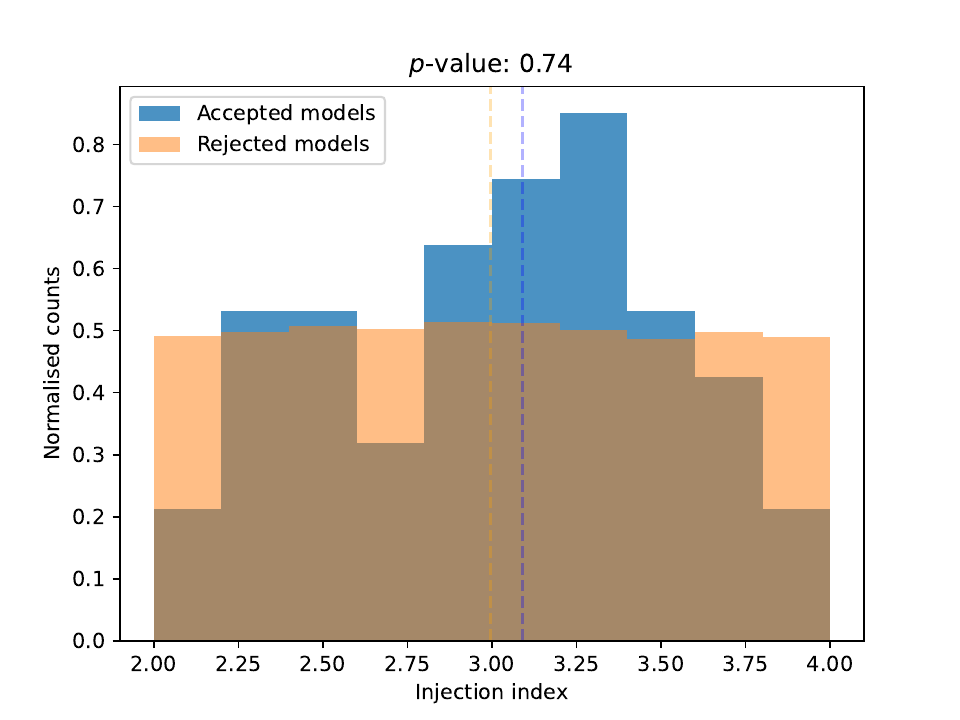}
    \caption{Distributions of model parameters based on the constrained spectra in Figure \ref{fig:plot_spec}, where blue histograms display models consistent with the radio and X-ray flux measurements and upper limits, and orange histograms display models that are not (i.e. predicting fluxes inconsistent with our measurements or higher than our radio upper limits). The $p-$values from two-sample Wilcoxon Mann-Whitney U tests are displayed above each plot (rounded to two decimal places -- the $p-$value for the $\gamma_{\text{max}}$ distributions is $\sim10^{-16}$). The medians of each distribution are shown by dashed lines. Note the skew of the Lorentz factor distributions is due to the requirement of $\gamma_{\text{min}}\leqslant\gamma_{\text{break}}\leqslant\gamma_{\text{max}}$ after drawing from log-uniform distributions.}
    \label{fig:hists}
\end{figure*}
In Figure \ref{fig:plot_spec} we display the synchrotron (blue) and inverse-Compton (orange) spectra of the models that are consistent with measurements at 144\,MHz and at 1\,keV (so that we capture models where predicted fluxes match our measurements within their errors), as well as VLA 3$\sigma$ upper-limits at 1.5\,GHz and 3\,GHz. We note that the archival data based on the legacy VLA system (except for the 3\,GHz and 15\,GHz data which is from the upgraded JVLA) has relatively poor sensitivity compared to LOFAR -- the snapshot data at 8\,GHz has a 3$\sigma$ upper limit ($\sim130\mu$Jy beam$^{-1}$) significantly higher than the 15\,GHz data and therefore its constraints to not affect the analysis. In fact, the only upper limit that constrains the spectra is from the 1.5\,GHz data, as clearly seen in the figure. We further note that the archival radio frequency data with the VLA at 5\,GHz and above lack the short baselines that LOFAR has (prohibiting imaging at 1.2\,arcsec resolution), and are therefore not used in the fitting (limits at 5\,GHz and above are plotted in Figure \ref{fig:plot_spec} for reference purposes). Figure \ref{fig:plot_spec} shows that the synchrotron and inverse-Compton losses are extremely strong beyond a few hundred MHz -- the diffuse lobe spectra predicted suggest that there is no flux density in the diffuse plasma above $10^{-14}$\,Jy at 15\,GHz. Even at 5\,GHz, the predicted flux density (blue curves) is $\ll1$\,$\mu$Jy, which is beyond the detection capabilities of current interferometers. The detection of the diffuse IC Ghost plasma in 3C\,294 requires observations at the lowest frequencies in the observational window.

In Figure \ref{fig:hists} we plot the distributions of the model parameters of those models that can be fitted to our observational constraints on the aged plasma ("accepted"; light blue) and those models that cannot ("rejected"; orange), to understand the difference in physical characteristics that can produce ghost IC emission. As a note, the majority of models are rejected (99.8\%), indicating the rarity of this plasma\footnote{a large number of models are not considered due to non-physical electron energies, caused by very low $\gamma$ electrons radiating in a low-strength magnetic field, which almost impossible to produce with this distribution of parameter sets.} under the assumption that the parameter range used produces lobe emission that can be detected. For each of the six free parameters in our analysis, we perform two-sample Wilcoxon Mann-Whitney U (WMW) tests between the distributions of accepted and rejected models -- the $p-$values indicating the probabilities of non-rejections for the null hypothesis that the medians of the distributions are similar are stated above each plot. The most significant differences are for the distributions of maximum Lorentz factors and the lobe magnetic field strengths: the diffuse emission seen in 3C\,294 can only be produced if $10^{2.0}\leqslant\gamma_{\text{max}}\leqslant10^{3.5}$ (top right panel) and if $10^{-9}\leqslant B \,\text(T) \leqslant10^{-8}$. In the case of the $B-$field, the $p-$value $\approx0.16$ is significantly higher than that required for a result to be statistically significant at a 95\% confidence interval -- however, clearly this is due to the medians being similar while the ranges of the distributions are different. A two-sample Kolmogorov-Smirnov (KS) test for the distributions of the accepted and rejected $B-$field values yields a $p-$value of $10^{-14}$, indicating a significant difference. 

\begin{figure}
    \centering
    \includegraphics[width=1\linewidth]{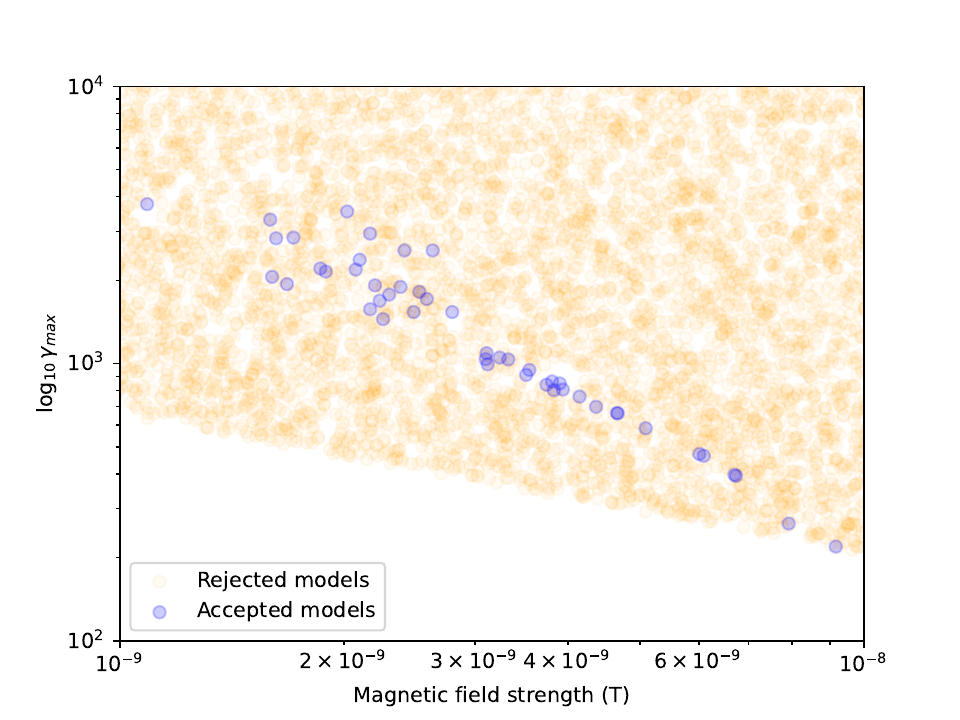}
    \caption{Lobe magnetic field strength against $\gamma_{\text{max}}$, for the models which fit (blue) and do not fit (orange) to our observational data (zoomed into the narrow range of the accepted models), as in Figure \ref{fig:hists}. A clear correlation between the two parameters is seen, reflecting the basic physics of synchrotron emission.}
    \label{fig:b-gmax}
\end{figure}

The median values (blue dashed lines) for each of the parameters that produce emission consistent with our data are; $\gamma_{\text{min}}=1.13$, $\gamma_{\text{max}}=1447.12$, $\gamma_{\text{break}}=2.98$, $B=2.80$\,nT, $p=3.08$, and  $\Delta p=1.36$ (tabulated in Table \ref{tab:model_params}). The most striking result is that the accepted models have low-frequency power-law spectra that break at very low frequencies -- electron populations exist with energies only in the range $1\leqslant\gamma\leqslant3$ before the energy spectrum breaks, in the diffuse lobes of 3C\,294. This corresponds to a break-frequency of the power-law of 275\,MHz (rest-frame), or 100\,MHz in the observed frame (i.e. it is probable that the emission we detect with LOFAR at 144\,MHz is already rapidly ageing), suggesting this critical frequency range to detect such plasma. Using the median fitted values in Table \ref{tab:model_params} (except for the $B-$field), we determine an equipartition magnetic field strength of $B_{\text{equip}}=2.5\times10^{-8}$\,T (250\,$\mu$G), and so while large, our fitted median value is $B\approx0.1B_{\text{equip}}$, typically lower than what is found for the other FR-II radio galaxy lobes where there are X-ray lobe detections of ICCMB emission \citep{cros05}. If the lobe magnetic field had been lower than $10^{-9}$\,T (as might be expected in the most aged plasma of very long-lived sources, e.g. \citealt{shul24}), we would not have detected the diffuse emission. We note that the constrained narrow ranges for $\gamma_{\text{max}}$ and $B$ are due to an association -- if there is a population of higher $\gamma$ electrons, you need a lower magnetic field strength to produce the same intensity of emission (with all other parameters fixed), as shown in Figure \ref{fig:b-gmax}.

As expected, the fits are insensitive to the injection index $p$ and the change in the power-law index after the break. The extreme curvature of the spectra constrained by our detections and limits seen in Figure \ref{fig:plot_spec} is largely due to the dependence of the emissivity on the magnetic field (as $B^{2}$), while the position of the curvature is determined by the Lorentz factors. Overall, the median total energy in the diffuse part of both lobes equals $\sim3\times10^{57}$\,J (or $\sim3\times10^{64}$\,erg), as determined by the \texttt{PYSYNCH} code. This is significantly greater than the thermal energies in the hot gases of even the most massive clusters ($M\sim10^{14}M_{\odot}$), even at extreme temperatures of 5\,keV. This suggests strong feedback effects\footnote{Of course, our energy estimates depend on a number of assumptions such as the volume of the regions being modelled, the volume filling factor, and the lack of non-radiating particles contributing to the total energy -- we do not expect order of magnitude differences between our assumptions and reality.}, given that simulations predict as much energy going into shock-heating of the environment by the lobes \citep[e.g.][]{hakr13}. As such gas properties are unlikely for the protocluster surrounding 3C\,294, the only reasonable explanation is that this energy is deposited into the protocluster over a long timescale. Using the formula for the synchrotron timescale (assuming continuous injection of particles):
\begin{equation}
    t_{\text{age}}=1590\frac{B^{\frac{1}{2}}}{B^2+B^2_{B_{\text{CMB}}}}\nu_{\text{break}}^{-\frac{1}{2}}\,\text{Myr},
\end{equation}
using the median $B-$field value of 2.80\,nT, the equivalent magnetic field strength of the CMB field of $B_{\text{CMB}}=2.5$\,nT, and assuming a break at $\gamma_{\text{break}}=3$ (equating to 275\,MHz in the rest-frame), this implies an age of $t_{\text{age}}\approx13$\,Myr. This is sensitive to $\gamma_{\text{break}}$ which is poorly constrained in our analysis -- if we chose $\gamma_{\text{break}}$ to be $10^0$ and $10^3$ as in the tail ends of the `accepted' distribution shown in the middle-left panel of Figure \ref{fig:hists}, then $t_{\text{age}}$ is $\sim50$\,Myr and $0.14$\,Myr, respectively.  

In general our results signify that ICCMB is indeed the physical mechanism driving the radio and X-ray emission as suggested by \citep{fabi01,fabi03}: the combination of low Lorentz factors ($\gamma_{\text{max}}\lesssim10^3$), a rapid break in the electron energy spectrum ($\gamma_{\text{break}}\gtrsim\gamma_{\text{min}}$), and magnetic field strengths down to an order of magnitude below the equipartition value give ample evidence for the lack of GHz-frequency radio emission (beyond significant detection with any other instrument at GHz frequencies).  
\subsection{Resolved spectral index mapping}
\label{sect:hotspots}
In this section, we analyse the resolved spectral index distributions in 3C\,294 in order to determine the energetics of the high-energy regions. The origin of the diffuse plasma giving rise to the ICCMB emission is suggested to be old plasma from past jet activity pointing in the NW-SE direction \citep{erlu06}. The question of whether the source is precessing, restarting, or both, can be directly probed by understanding how and where the jets had terminated in the past, using our broad-band radio data. 

The jet termination process can be distinguished between models that describe all hotspots being connected to the jet \citep[e.g. the `splatter-spot' model;][]{will85} and those that are disconnected \citep[e.g. in the case of a precessing jet;][]{cox91}. In the former case, all hotspots will have similarly flat spectral indices if they are all connected to the current jet in a continuous flow. Meanwhile, the lobe regions at a large physical distance from the primary hotspots can be characterised as left-over plasma as the jet precesses; in this case, one would expect to see a smooth steepening of the spectral index along the lateral extension near the edge of the source. This allows us to conclude the origin of the diffuse NW (and SE) plasma responsible for the ICCMB emission. 

We obtain total intensity maps at 1.2\,arcsecond resolution (convolved with a circular beam of this size) using our 144\,MHz, 1.4\,GHz, and 3\,GHz data (consistent in the $uv$-plane by applying Gaussian tapering during imaging -- the same maps used to measure flux densities in Section \ref{sect:modelling}), and correcting any positional offsets by fitting the northern outer hotspot with a Gaussian across the three images, and aligning the images using the mean position of the fitted Gaussians. We apply the same method to produce 0.3\,arcsecond maps, using our 144\,MHz, 5\,GHz, 8.4\,GHz, and 15\,GHz maps. With all spectral index measurements, we quote statistical (propagation of RMS noise values at each frequency) errors. Spectral index maps are made using the \texttt{BRATS} software \citep{harwood13}, which, for all pixels, performs a regression fit of a single power-law across the band. For spectral ageing maps, radiative losses are directly calculated given an input magnetic field strength and fitted to the observed spectra to determine the break frequency on a pixel-by-pixel basis (see \citealt{harwood13} for more details). 
\subsubsection{The lobes}
\begin{figure}
    \centering
    \includegraphics[width=1.1\linewidth,trim={5cm 0.3cm 0cm 0.2cm},clip]{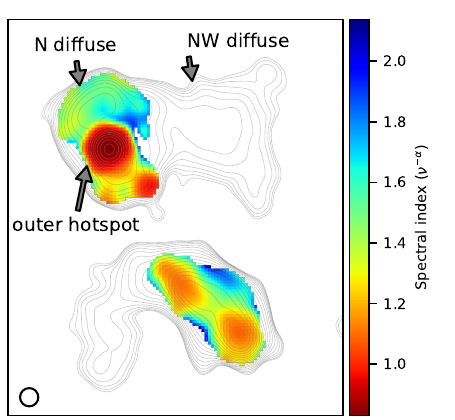}
    \includegraphics[width=1.1\linewidth,trim={5cm 0.3cm 0cm 0.2cm},clip]{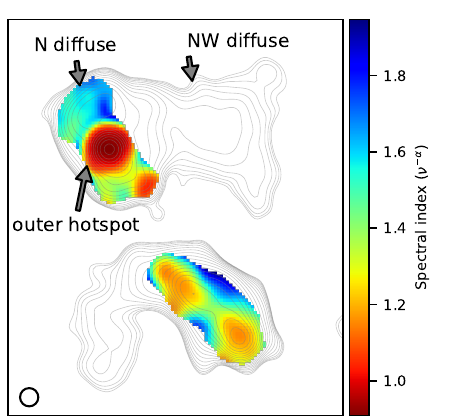}
    \caption{Spectral index maps at a common angular resolution of 1.2\,arcsec using the 144\,MHz and 1400\,MHz maps (top) and also including the 3000\,MHz map (below). Beam size of 1.2\,arcsec is shown on the lower left.}
    \label{fig:1asec_spix}
\end{figure}

In Figure \ref{fig:1asec_spix} we display two and three-point spectral index maps at 1.2\,arcsec angular resolution, overlaid with the 144\,MHz contours to display the full extent of the diffuse plasma. In the top panel, we omit the 3\,GHz data in order to display the spectral index with more regions of flux detected, particularly in the`N diffuse' region. Notably, Figure \ref{fig:1asec_spix} shows that the `N diffuse' plasma north of the northern outer hotspot displays a mild gradient in the transverse direction (steepening towards the west). The spectral indices in the top panel (without the 3\,GHz data) steepen from a value of $\alpha^{144}_{1400}\approx1.42\pm0.03$ to $1.53\pm0.04$ towards the diffuse region, and show a similar behaviour in the bottom panel, which includes the 3\,GHz data. This suggests the history of the plasma flowing from east to west, after initial particle acceleration in the northern lobe, and is consistent with jet precession models. The entire northern lobe emission beyond the northern outer hotspot is disconnected from any ongoing jet activity: the sharp spectral index discontinuity between the outer northern hotspot ($\alpha^{144}_{3000}\gtrsim0.91\pm0.0005$) and the `N diffuse' region ($\alpha^{144}_{3000}\gtrsim1.51\pm0.13$) in the bottom panel imply aged plasma from an older episode of jet activity, which has since progressed due to adiabatic expansion, rather than implying plasma protruding from the hotspot (as in the case of plumed FR-I objects, in which case a smooth spectral steepening is common). An identical analysis has been made for the famously restarting source 3C\,388 \citep{roet94,brie20}, where its eastern lobe is similar in morphology and spectral index to the northern lobe of 3C\,294.

\subsubsection{The jet and hotspots}
While we will present a broadband analysis of the radio-X-ray emission of the jet, jet knots, and hotspots in a forthcoming paper, we present here an analysis of the spectral ageing in the hotspots to ascertain whether the inner compact hotspots represent restarted activity.

\begin{figure}
    \centering
    \includegraphics[width=1.1\linewidth,trim={5.3cm 0.15cm 0cm 0.2cm},clip]{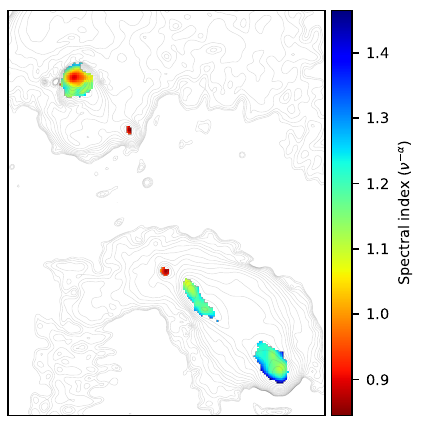}
   \caption{Spectral index map at a common angular resolution of 0.3\,arcsecond using the 144\,MHz, 5\,GHz, 8.4\,GHz, and 15\,GHz maps, fitted with a single power-law. Contours describe the 0.3\,arcsecond 144\,MHz data.}
   \label{fig:03asec_spix}
\end{figure}

\begin{figure}
    \centering
    \includegraphics[width=1.1\linewidth,trim={5.3cm 0.15cm 0cm 0.2cm},clip]{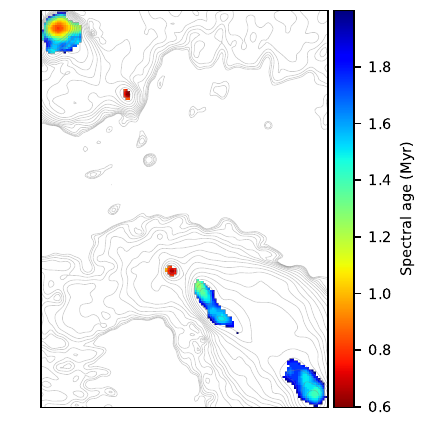}
   \caption{Spectral age map at a common angular resolution of 0.3 arcsecond using the 144\,MHz, 5\,GHz, 8.4\,GHz, and 15\,GHz maps. Contours describe the 0.3\,arcsecond 144\,MHz data.}
   \label{fig:03asec_specage}
\end{figure}

In Figure \ref{fig:03asec_spix} we display a spectral index map at 0.3\,arcsecond resolution using the 144\,MHz, 5\,GHz, 8.4\,GHz, and 15\,GHz data. For the northern hotspot pair, the inner hotspot has $\alpha_{15000}^{144}\approx0.85\pm0.09$, consistent with the outer hotspot having $\alpha_{15000}^{144}\approx0.90\pm0.001$. CConversely,for the southern side, the inner hotspot with $\alpha_{15000}^{144}\approx0.86\pm0.06$ is flatter than the jet structure ($\alpha_{15000}^{144}\approx1.12\pm0.048$) and the outer hotspot ($\alpha_{15000}^{144}\approx1.11\pm0.019$). Therefore, given the larger statistical errors in the northern inner hotspot due to a few number of pixels, the inner hotspots likely represent a new phase of jet activity. The detected southern jet and both outer hotspots are likely remnants of past activity. We use the \texttt{BRATS} software to ascertain the spectral ages of the hotspots (shown in Figure \ref{fig:03asec_specage}). We fit to the data models where the magnetic field is at equipartition, as typical for hotspots \citep[e.g.][]{harr00,hard01} -- this is $\sim10$\,nT as measured for 3C\,294 -- the inner and outer hotspots have similar equipartition values), specifying the Jaffe-Perola \citep{jpmodel} model. Kardashev-Pacholczyk \citep[KP;][]{kard62,pach70} and Tribble \citep{tribblemodel} models were also fitted, with similar results to the JP model. The JP and KP models differ in their treatment of the pitch angle between the particle velocity and magnetic field vectors, while the Tribble model is similar to the JP model but with a Gaussian random field. Given that our results do not distinguish between the three, we refer the reader to a more detailed explanation and a comparison of these models presented by \citep{harw17_specmodels}. Figure \ref{fig:03asec_specage} shows that the inner hotspots have an age $\lesssim0.6$\,Myr, while the outer northern hotspot has had $\sim 1$\,Myr since acceleration, consistent with the short remnant phase expected \citep[e.g.][]{maha18,turn20_xray} and that restarted jets must merge with the pre-existing plasma on a shorter timescale than the previous jets \citep[see review by][]{maha23_review}. Further data points, particularly at 1.5\,GHz at sub-arcsecond resolution, are needed to affirm these results with higher significance. In summary, our analysis gives some evidence that the jets in 3C\,294 are both precessing and restarting, with at least three generations of jet activity uncovered in our radio maps.

%\begin{figure}
%    \centering
%    \includegraphics[width=1.1\linewidth,trim={5.3cm 0.4cm 0cm 0.2cm},clip]{plots/3C294_spix_03asec_144-5000-8400.pdf.pdf}
%    \includegraphics[width=1.1\linewidth,trim={5.3cm 0.4cm 0cm 0.2cm},clip]{plots/3C294_spix_03asec_5000-8400.pdf.pdf}
%    \caption{Spectral index maps at a common angular resolution of 0.3 arcsecond using the 144\,MHz, 5000\,MHz and 8400\,MHz maps (top) and the same without the 144\,MHz map (below). The 15000\,MHz data are not included due to a significant deviation from a power-law driving large errors. }
%    \label{fig:03asec_spix}
%\end{figure}
\section{Discussion}
\label{sect:discussion}
\subsection{Lobe physical conditions}
\cite{fabi03} and \cite{erlu06} suggested that the most energetically plausible explanation for the presence of the diffuse X-ray emission in 3C\,294 is through ICCMB of the relativistic electron populations in the lobes. Given their X-ray-only constraints, and an assumption of a dominant $\gamma\sim10^3$ electron population to produce the 1\,keV X-rays, they determined a total energy in relativistic electrons to be $\sim5\times10^{59}$\,erg. This is a strict lower limit to the total energy content in the lobes, as they assumed a monochromatic electron energy distribution (in the absence of a radio detection to invoke a parametrised power-law distribution).

Our new constraints on the physical properties of the diffuse emission in IC Ghosts have given new insights: contrary to studies suggesting a high low-energy cutoff in the lobes \citep[i.e. $\gamma_{\text{min}}\sim10^3$][]{erlu06,fabi09}, our analysis suggests much lower energies -- the most probable $\gamma_{\text{min}}$ is 1 (Figure \ref{fig:hists}), breaking very quickly at $\gamma_{\text{break}}\approx3$. This suggests that the radio lobes in these circumstances store far more energy than typically assumed. To compare with previous results, \citep{erlu06} calculated that the energy stored in electrons in the diffuse region (based on the X-ray luminosity) is $E>3.91\pm1.42\times10^{59}$\,erg based on a monochromatic electron energy distribution. Using our constraints, we find a median value of $E\approx3\times10^{64}$\,erg (in both lobes together; Section \ref{sect:modelling}). We also classify this as a lower limit -- the fraction of non-radiating particles (through the required entrainment by the lobes; e.g. \citealt{bick84}) is unknown, but they contribute to the energy budget.

Clearly, the energetics in 3C\,294 are extreme compared with those determined for the lobes of other FR-II radio galaxies \citep[e.g.][]{cros18}. The relatively high value for the magnetic field strength could be driven by compression of the plasma by surrounding turbulence (as expected in protoclusters); however, the X-ray data seemingly do not predominantly represent hot gas from the ambient medium \citep[][]{fabi03}. Deeper X-ray data are needed to ascertain the nature of any radio lobe plasma interaction with its hot gas surroundings. Otherwise, our analysis suggests that 3C\,294 represents an extremely high power source near the quasar epoch, with a jet kinetic power of $Q_{\text{jet}}=\frac{3\times10^{64}\,\text{erg}}{13\,\text{Myr}}\approx7\times10^{42}$\,W. The luminosity of the core, a proxy for the power output of the AGN, is $1.1\times10^{38}$\,W (based on the X-ray spectra; \citealt{fabi03}). The orders of magnitude difference between these quantities reflects different physical origins -- the jet kinetic power as measured with radio observations is largely a function of the particle acceleration and radiative loss mechanisms over Myr timescales, while the X-ray power of the core may trace accretion properties over much shorter timescales \citep[e.g.][]{sobo09}.
\subsection{Hotspots with active particle acceleration?}
The injection index $p$, directly related to the slope of the power-law energy distribution, is traditionally assumed to be valued at 2 (or a spectral index $\alpha$ of 0.5) based on theoretical considerations of particle acceleration \citep{blan78}. However, more recent findings \citep[e.g.][]{harwood15} suggest that the injection index can be steeper, throwing into question what physical process is being measured at the locations of radio hotspots. Given the steep spectral indices of the inner hotspots ($\alpha\sim0.9$), it is interesting to discuss what components they represent -- hotspots (as traditionally known) represent the interaction between jet material and slow-moving lobe material, producing non-relativistic shocks through the Fermi-I process \citep{blan74}. The symmetry of the system suggests that the inner hotspots cannot be jet knots, as argued in Section \ref{sect:data}. It is possible that, due to the supposed precessionary nature of the source, the surrounding plasma is in a turbulent state where particles travelling up the jet cannot efficiently cross the shock front(s) before leaving the shock region, causing a low number of high-$\gamma$ particles to remain in the system. The most likely scenario is that since the restarted jets drive through plasma excavated by the previous phase, and therefore the ambient environment is less dense, the jet termination shock is weaker (possibly representing relativistic shocks, as in Fermi-II models). A forthcoming study measuring the broadband properties, including X-rays, will constrain the particle acceleration model. 
\subsection{Precession origin}
Our work, along with previous studies on 3C\,294 \citep[e.g.][]{fabi03,erlu06} wholly support jet precession as a mechanism to produce the transverse extent of the radio and X-ray emission. \cite{erlu06} suggest a precession timescale of $>10$\,Myr based on the X-ray emission alone. Given that the jets must precess on timescales longer than the age of the diffuse plasma (since the jets quickly merge with any pre-existing plasma in its path), we can constrain the precession timescale to be $>50$\,Myr (as implied by the lowest values of $\gamma_{\text{break}}$ in our modelling in Section \ref{sect:modelling}). Using this limit, and given the approximate transverse width of the radio emission detected by LOFAR ($\sim$12.7\,arcsec=109.2\,kpc), we find a upper limit projected lateral (transverse) velocity of the jet head due to precession of $v_{\text{jet,precession}}<2.13\times10^{3}$\,km\,s$^{-1}$. This value implies slower precession than that implied by models of precession originating from the Lens-Thirring or binary black hole interactions \citep{fell23}, in which case the origin of the precession could be due to galaxy-galaxy interactions as expected in the proto-cluster environment of 3C\,294. If $\gamma_{\text{break}}\approx 10^3$ (i.e. the age of the plasma is 0.14\,Myr), this puts an unphysical upper limit precession velocity at $v_{\text{jet,precession}}>c$, which does not provide further constraints. Deeper observations at 1.5\,GHz (i.e. with the upgraded JVLA, with an order of magnitude better sensitivity than the available data) will allow us to constrain the precession mechanisms more precisely.

Other than precession, an alternative possibility is that turbulence in the protocluster environment acts to displace the plasma of an otherwise straight (non-precessing) jet. Such hydrodynamical arguments have been used to describe the unusual morphology of systems such as NGC\,236 \cite{hard19_ngc326}, and of the longevity of the low surface brightness plasma of the radio source in Abell\,1033 \citep{dega17}. The lack of asymmetry in 3C\,294, unlike in the aforementioned objects, suggests this interaction may not be the primary mechanism for producing the lobe structure seen at 144\,MHz. Although unlikely, the random nature of the gas motions are expected have velocities of the order $10^2$\,km\,$^{-1}$ over $\lesssim100$\,kpc scales \citep{dima23} which is similar to the projected lobe width, and so we cannot rule out this plays a role in displacing the radio lobe plasma.
\subsection{Outlook}
This study has shown that low-frequency high dynamic range imaging is vital for the detection of the lobes of high-redshift radio galaxies (HzRGs). If 3C\,294 is representative of the population at the quasar epoch and earlier, sensitivity to $\gamma\sim1-10$ particles means that the detection of the aged lobe material in the population requires as low a frequency as possible, while we would only be sensitive to lobes containing magnetic field strengths above a few nT. High-dynamic range and high resolution ($\sim1$\,arcsec) imaging with the LOFAR Low Band Antennas (LBA) with a central frequency of 50\,MHz has shown to be possible \citep{groe23}, and improvement of the calibration and imaging capabilities in this regime is vital for the detection of all HzRG lobe material at a fixed surface brightness limit. Future surveys from the upcoming LOFAR2.0 \citep{lofar2.0}, utilising simultaneous LBA and HBA observing, will allow strong conclusions on the properties of these objects while more are being uncovered with the current LOFAR system at 144\,MHz (e.g. HDF\,130; Mahatma+ in prep). The future SKA-LOW telescope, while planned to provide up to an order of magnitude improvement in raw sensitivity over LOFAR, has much poorer angular resolution and therefore radio galaxies smaller than $\sim20$\,arcsec (the angular size of 3C\,294) will only partially be resolved. Therefore, future searches for the IC Ghost population to the highest redshifts are most promising using SKA-LOW VLBI (as envisioned by the Low-frequency Australian Megametre-Baseline Demonstrator Array\footnote{\url{https://www.atnf.csiro.au/projects/instrumentation/lambda/}} project, LAMBDA). With more complete low-frequency surveys down to micro-Jansky sensitivity, we may uncover a population of HzRGs previously missed. 
\section{Conclusions}
This study describes the first measurement of the radio emission associated with diffuse X-ray Ghost plasma, systems which have evaded classification in the past, and the number density of which severely underpredicts models. Using the radio and X-ray observational constraints of the diffuse plasma of the 3C\,294 lobes, we determine that:
\begin{itemize}
    \item The 3C\,294 lobes are dominated by a low-energy electron population with a steep electron spectrum, so that high redshift radio lobes in general, if detected,  probe very low $\gamma$ particles.
    \item The magnetic field strength between 1-10\,nT and the dominance of inverse-Compton losses due to the CMB in the lobes are central to explaining the X-ray emission and the electron energy distribution.
    \item The synchrotron timescale yielded by the allowable Lorentz factors is 13\,Myr, consistent with typical lobe lifetimes \citep{pinj25} and precession timescales \citep[e.g.][]{krau19,misr25}.
    \item The detectability of ICCMB in the lobes of 3C\,294 is not related to the particle injection index, and by extension, therefore, to the efficiency of particle acceleration. Wide-area low-frequency surveys should be sensitive to many such objects with a wide distribution of radio luminosities.
    \item There are strong indications that 3C\,294 has jets which are both precessing and restarting, while restarting on very short ($<1$\,Myr) timescales.
\end{itemize}
\label{sect:conclusions}
This study motivates wide area searches for ICCMB plasma from a number of bright radio galaxies with existing deep X-ray data. With systematically selected samples, driven by the near-full-northern sky coverage of LoTSS DR3 (Shimwell et al., in review), we will soon be able to determine better constraints on the fraction of ICCMB ghosts and the detectability of such objects with future sensitive low-frequency arrays.
\section*{Acknowledgements}
We thank the anonymous referee for helpful comments in improving this manuscript.

LKM is grateful for support from a UKRI FLF [MR/Y020405/1] and STFC support of LOFAR-UK [ST/V002406/1].

This research made use of the University of Hertfordshire
high-performance computing facility and the LOFAR-UK computing
facility located at the University of Hertfordshire (\url{https://uhhpc.herts.ac.uk}) and supported by
STFC [ST/P000096/1].

LOFAR is the Low Frequency Array, designed and constructed by ASTRON. It has observing, data processing, and data storage facilities in several countries, which are owned by various parties (each with their own funding sources), and which are collectively operated by the ILT foundation under a joint scientific policy. The ILT resources have benefited from the following recent major funding sources: CNRS-INSU, Observatoire de Paris and Université d'Orléans, France; BMBF, MIWF-NRW, MPG, Germany; Science Foundation Ireland (SFI), Department of Business, Enterprise and Innovation (DBEI), Ireland; NWO, The Netherlands; The Science and Technology Facilities Council, UK; Ministry of Science and Higher Education, Poland; The Istituto Nazionale di Astrofisica (INAF), Italy.

The National Radio Astronomy Observatory is a facility of the National Science Foundation operated under cooperative agreement by Associated Universities, Inc.

This research made use of APLpy, an open-source plotting package for Python \citep{aplpy2012,aplpy2019}.
%%%%%%%%%%%%%%%%%%%%%%%%%%%%%%%%%%%%%%%%%%%%%%%%%%
\section*{Data Availability}
The data underlying this article will be shared on reasonable request to the corresponding author.



\bibliographystyle{mnras}
\bibliography{references} % if your bibtex file is called example.bib

%\begin{thebibliography}{99}
%\end{thebibliography}

%%%%%%%%%%%%%%%%%%%%%%%%%%%%%%%%%%%%%%%%%%%%%%%%%%

%%%%%%%%%%%%%%%%% APPENDICES %%%%%%%%%%%%%%%%%%%%%

%\begin{figure*}
%    \centering
%    \begin{tabular}{cc}
%    \includegraphics[height=5cm, trim=0.5cm 1cm 0.3cm 0cm, clip]{plots/3c34_falsecolor.png}     &    \includegraphics[height=5cm, trim=0.5cm 0cm 0.3cm 0.5cm, clip]{plots/3c320_falsecolor.png} \\
%    \end{tabular}
%    \includegraphics{}
%    \caption{Caption}
%    \label{fig:my_label}
%\end{figure*}

%%%%%%%%%%%%%%%%%%%%%%%%%%%%%%%%%%%%%%%%%%%%%%%%%%

% Don't change these lines
\bsp	% typesetting comment
\label{lastpage}
\end{document}